# SEED-APPLIED BIOCONTROL: TOWARDS A NEW GENERATION OF PROTECTIVE STRATEGIES

T. FRANÇOIS[(1)], D. SOURDEVAL[(1)], R. LETAC[(1)], S. BOUTET[(1)], C. BROSSE[(1)], B. COLLET[(1)], C-E. KOUTOUAN[(2)], C. SACCARAM, A. ROYER[(1)], A. BENAMAR[(1)], R. MIRAY[(3)], E. FAUCHER[(3)], F. CHAUFFOUR[(3)], J-L. CACAS[(1)], M. CORSO[(1)], L. RAJJOU[(1,3)]

[(1)] Université Paris-Saclay, INRAE, AgroParisTech, Institute Jean-Pierre Bourgin for Plant Sciences (IJPB), 78000, Versailles, France.
[(2)] FNAMS, Fédération Nationale des Agriculteurs Multiplicateurs de Semences, Impasse du Verger - Brain-sur-l'Authion - 49800 Loire-Authion
[(3)] SEED IN TECH, Centre INRAE, route de St-Cyr, 78026 Versailles Cedex

**ABSTRACT**
Seeds are fundamental to agricultural productivity but also act as potential vectors for pathogens, leading to substantial losses at seedling emergence. This study focuses on one major arable crop (i.e., Wheat) which is subject to significant biotic stress. The objective is to evaluate the efficacy of biocontrol agents as sustainable alternatives to conventional seed treatments, with the goal of enhancing seed protection and increasing tolerance to biotic threats. Multi-season trials were conducted, integrating physiological, molecular, and metabolic analyses to elucidate the responses of treated seeds across key developmental stages (development, maturation, and germination). The findings indicate that the effectiveness of biocontrol agents is influenced by both the genotype and physiological stage of the seeds. Biocontrol treatments were shown to induce specific defence mechanisms and reduce pathogen pressure. This research study will pave the way to develop robust assessment tools for evaluating the performance of biocontrol strategies and to optimise their deployment in seed protection frameworks.

**RÉSUMÉ**
Les semences, essentielles à la productivité agricole, sont également des vecteurs d'agents pathogènes susceptibles d'entraîner des pertes importantes à la levée. Nos travaux portent sur une espèce de grande culture (blé tendre), d'intérêt économique majeur et exposée à de fortes pressions biotiques. L'objectif est d'évaluer l'efficacité d'agents de biocontrôle en tant qu'alternative aux traitements conventionnels pour protéger les semences et renforcer leur tolérance aux bioagresseurs. Les essais, conduits sur plusieurs campagnes de culture, intègrent des analyses physiologiques, moléculaires et métaboliques pour décrypter la réponse des semences traitées à différents stades (développement, maturation, germination). Les résultats montrent que l'efficacité des agents de biocontrôle dépend du génotype et du stade physiologique des semences. Les solutions de biocontrôle induisent des mécanismes de défense spécifiques et limitent la pression des agents pathogènes. À terme, cette recherche vise à développer des outils robustes pour évaluer l'efficacité des biosolutions et optimiser leur utilisation en protection des semences.

**Introduction**

Modern agricultural production relies heavily on seed quality, which determines both the genetic potential and field performance of crops (Rajjou et al., 2012; Finch-Savage & Bassel, 2016; Reed et al., 2022). However, beyond their essential role in reproduction, seeds also act as vectors for the transmission of various pathogens and pests, which can lead to significant economic losses (Baker & Smith, 1966; Elmer, 2001; Denancé & Grimault, 2022; Hiddink et al., 2023). Conventional seed treatments, typically based on synthetic fungicides and insecticides, are increasingly being called into question due to their negative impacts on the environment, human health, and the development of resistance. Consequently, the optimisation and development of seed-applied biocontrol agents represent an innovative approach, with the potential not only to enhance seedling emergence but also to trigger effective early defence responses against biotic stressors (Lamichhane et al., 2022). Recent research has focused on the integration of microorganisms into seed coating formulations, enabling controlled release of active compounds and prolonged protection during storage and germination (Paravar et al., 2023; Fan et al., 2024; Köhl et al., 2024). Seed-applied biocontrol agents may act through a variety of mechanisms, which broadly fall into two categories: direct inhibition of pathogens and the induction of systemic defence responses in the embryo and emerging seedling (Song et al., 2017). The concept of "*seed defence biopriming*" enables the priming of seeds for enhanced immune responses without compromising growth, an advantage over certain chemical treatments that often entail a growth-defence trade-off during early development. However, seed protection should not be confined solely to post-harvest interventions such as coating, film coating, or biopriming. A more integrated approach involves acting earlier in the seed production process by targeting the seed-producing mother plant itself. Applying biocontrol agents to mother plants could reduce vertical transmission of pathogens and thus improve the production of healthy seed lots. This paradigm shift, considering seed health protection from the stage of seed development, opens up new perspectives for strengthening the sustainability of seed systems while reducing reliance on curative treatments. Currently, data on the protective effects of seed-applied biocontrol agents remain limited, both in terms of efficacy and the underlying biological processes involved. In particular, further investigation is needed into the physiological, molecular, and metabolic changes contributing to protection against pathogens. It remains to be determined to what extent biocontrol agents alter gene expression, metabolism, or the seed-associated microbiota, and how these changes impact the ability of seedlings to resist pathogens during early development. Moreover, the role of plant genotype in responsiveness to these treatments is still poorly understood, as is the stability and heritability of any observed benefits across successive seed generations. These knowledge gaps currently hinder the broader adoption of such technologies in the seed industry and conventional agriculture. The research presented here was designed to address these uncertainties and to advance understanding of the effects of seed-applied biocontrol agents. It aims to characterise, in an integrated manner, the responses induced by such agents when applied either to the seed-producing mother plant or directly to the seed *via* coating. Our study combines molecular analyses (transcriptomics, metabolomics) and phenotypic evaluations (germination, vigour). Our work focuses on a single agronomically important model species, bread wheat (*Triticum aestivum*), cultivated under conditions representative of seed production systems. The main hypotheses tested are as follows: (1) application of a biocontrol agent, either to the mother plant during seed development or directly to post-harvest seed, could induce measurable defence responses at various biological levels; (2) these responses may vary according to the genotype; and (3) the effectiveness of protection could be correlated with identifiable molecular changes. This article presents a selection of results obtained from the experiments conducted in the project. These findings allow for discussion not only of the efficacy of the tested treatments, but also of the mechanisms involved and the potential for application within seed production systems.

**Materials and Methods**

*Application of treatments:* This study was conducted on a single agronomically important model species, bread wheat (*Triticum aestivum* L.), grown under conditions representative of seed production systems. Four genotypes (Genotype 1, Genotype 2, Genotype 3, and Genotype 4) were selected based on their genetic diversity and varying susceptibility to biotic stressors, in order to assess genotype-dependent responses to treatments. For treatments applied to mother plants, both biocontrol agents and plant resistance inducers (PRI) were used. Among the compounds applied were acibenzolar-S-methyl (ASM), known for inducing systemic acquired resistance, and copper (Co), used both as a defence elicitor and as a reference fungicide. In addition, the beneficial microorganism *Pseudomonas chlororaphis* was included as biocontrol (BC) agent, as well as a control with water (W) treatment. These treatments were applied *via* foliar spray at different phenological stages: heading, flowering/anthesis, grain development, and seed maturation. Seed samples were collected at early maturation (S1) and late maturation (S2), with manual dissection of the embryo (E) and endosperm + seed coat (ESC), followed by immediate freezing in liquid nitrogen. Mature seeds (M) and germinating seeds (G) were also collected, cleaned, dissected, and prepared for subsequent analyses. For post-harvest treatments, dry seeds were film-coated with solutions containing the same active compounds or beneficial microorganism, in order to compare the effects of direct seed applications with those of maternal plant treatments. The coated seeds were dried under controlled temperature and stored under standard short-term conditions. They were subsequently germinated, and the germinated seeds were dissected to isolate the root (R) and shoot (S) tissues of seedlings at 7- and 10-days post-germination (7D, 10D), which were then frozen at −80 °C.

*Germination tests:* To assess physiological and phenotypic responses, germination tests were conducted in accordance with the International Seed Testing Association (ISTA) standards (https://www.seedtest.org/), including measurements of germination rate and seedling vigour. Experiments were performed under controlled conditions with four biological replicates.

*Molecular analyses*: Molecular analyses, including transcriptomics and metabolomics, were conducted on Genotype 4.

*Transcriptomic analyses:* For transcriptomic analysis, total RNA was extracted from various tissues (embryo [E], *endosperm plus bran* [A], root [R], and shoot [S]) collected at different developmental (early maturation [S1], late maturation [S2], and mature seed [M]), germination (G), and seedling stages (7D and 10D). A total of 144 samples were analysed, corresponding to 2 tissue types × 6 developmental stages × 4 treatments × 3 biological replicates. The samples were ground in liquid nitrogen into a fine powder. A 20 mg portion of each sample was lysed at 65 °C in a CTAB-PVP extraction buffer (2% CTAB, 2% PVP, NaCl, EDTA, and β-mercaptoethanol) to neutralise polyphenols and polysaccharides, followed by sequential extraction with chloroform/isoamyl alcohol or phenol/chloroform/isoamyl alcohol (Kiss et al., 2024). The RNA was precipitated using isopropanol, washed with 75% ethanol, and resuspended in RNase-free water. RNA purity and integrity were verified using $A_{260}/A_{280}$ and $A_{260}/A_{230}$ ratios (within 1.8–2.2) and RNA Integrity Numbers (RIN ≥ 7), confirming suitability for transcriptomic analyses. RNA sequencing was performed using Next-Generation Sequencing (NGS) on the BGISEQ-500 platform. Obtained data were analysed to identify genes that were differentially expressed in response to the various treatments.

*Metabolomic analyses*: In parallel, untargeted metabolomic profiling was carried out on the same tissue samples using liquid chromatography–mass spectrometry (LC-MS/MS) to detect and quantify specialised metabolites potentially involved in induced defence mechanisms (Boutet et al., 2022). Metabolites were extracted from 20 mg of isolated tissue using a MeOH:MTBE:$H_2O$ buffer (1:3:1), separating polar/semi-polar fractions from lipids and proteins. Analyses were conducted using UHPLC–QToF under both positive and negative ionisation modes. Data were processed using MZmine 2.52, and metabolite annotation was performed using an in-house library, public databases (GNPS, MassBank, NIST), molecular networking, and SIRIUS software. Internal normalisation was done using rhamnetin. Metabolomic profiles were analysed to identify metabolites showing statistically significant changes in abundance across the different treatments.

*Experimental design and statistical analysis*: All experiments were organised using a factorial design, accounting for the variables: genotype, treatment type (maternal application vs. seed coating), and biological replication. Data were subjected to multivariate statistical analyses (p-value: 0.05), including ANOVA, multivariate variance analysis, and correlation analyses, to determine main effects and significant interactions.

**Results**

The full set of analyses conducted in this study enabled the evaluation of the effects of a biocontrol agent (*Pseudomonas chlororaphis*) applied either to the mother plant during seed production or directly to the seeds *via* film coating, in bread wheat (*Triticum aestivum* L.).

*Phenotypic analysis*: At the phenotypic level, no significant differences were observed in germination parameters or yield. Maximum germination rates reached 100% under all conditions, and the speed of seedling emergence remained comparable between treated and untreated seeds, regardless of the treatment method (maternal plant application or direct seed coating) or genotype (Genotype 1, Genotype 2, Genotype 3, Genotype 4). However, seeds coated with copper exhibited visible growth defects 10 days post-sowing. In contrast, a positive effect was observed on seedling vigour: genotypes Genotype 1 and Genotype 4 displayed a noticeably greater shoot height at 10 days after sowing when seeds were coated with *Pseudomonas chlororaphis*, without any negative impact on germination. These findings suggest that biocontrol treatments do not induce stress or growth/defence trade-offs during early development and may even enhance seedling vigour in certain genotypes.

Figure 1: Principal Component Analysis (PCA) of Transcriptomic Profiles in Wheat Seed and Seedling Tissues Under Biocontrol and Chemical Treatments

Figure 1 : Analyse en Composantes Principales (ACP) des Profils Transcriptomiques dans les Tissus de Grains et de Plantules de Blé en Réponse au Traitement de Biocontrôle et aux Traitements Chimiques

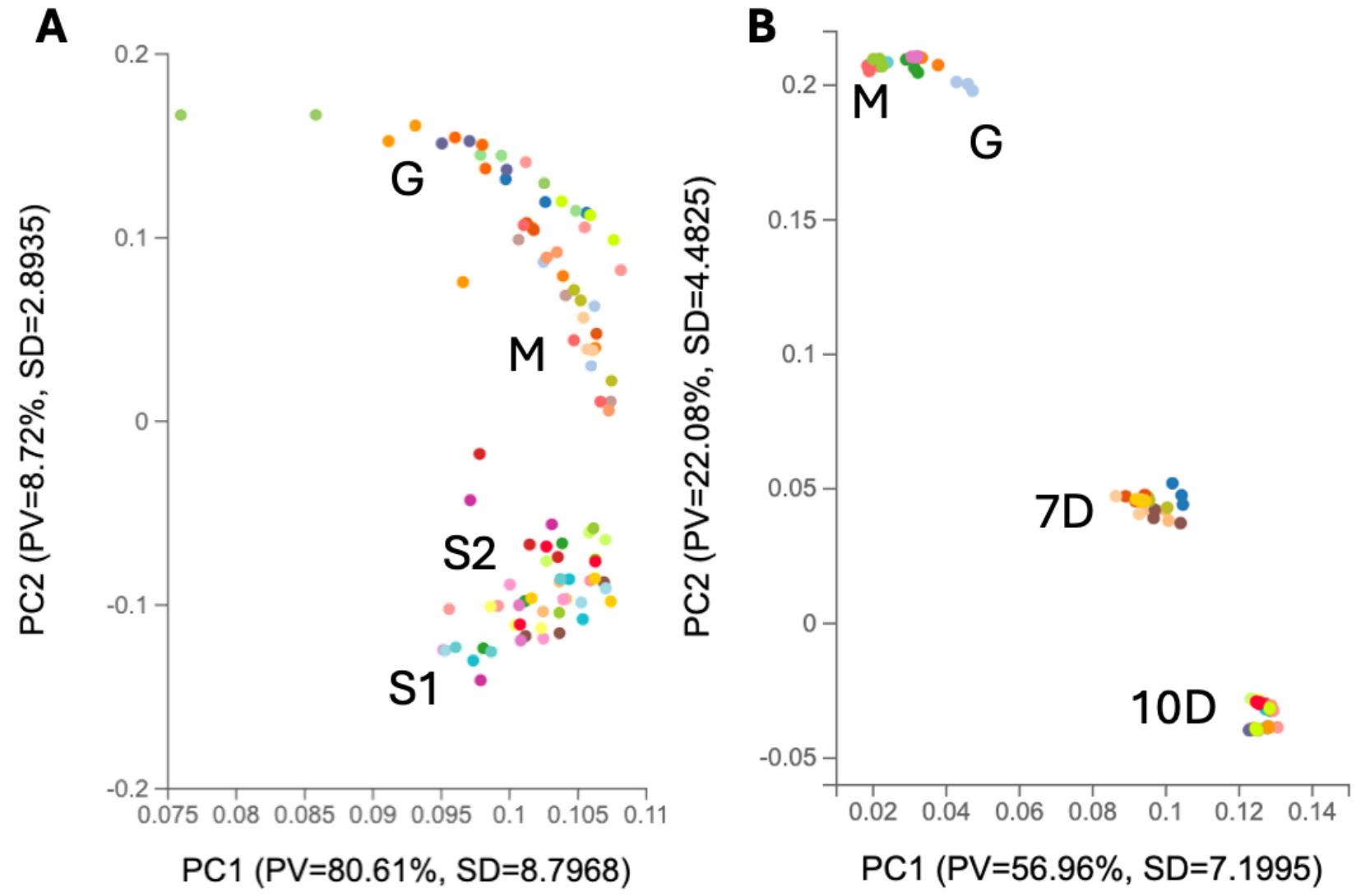

*Principal Component Analysis (PCA) of transcriptomic profiles from 144 RNA-seq samples collected across four tissue types, embryo (E), endosperm plus bran (A), root (R), and shoot (S), at six developmental stages: seed development (S1, S2, M), germinated (G), and early seedling growth (7 days [7D] and 10 days [10D] post-sowing). Samples were collected under four treatments: water*

*(control, W), acibenzolar-S-methyl (ASM), copper (Co), and the biocontrol agent Pseudomonas chlororaphis (BC), with three biological replicates per condition.* ***(Panel A)*** *PCA of samples from seeds produced on mother plants that received foliar treatments during seed development (pre-harvest). Principal Component 1 (PC1) explains 80.61% of total variance and separates samples by developmental stage, while PC2 (8.72%) captures additional variation among stages. Clear clustering is observed among seed maturation (S1, S2, M) and germinated (G) stages, indicating transcriptional reprogramming during development and germination.* ***(Panel B)*** *PCA of samples from seeds that received treatment via post-harvest seed coating. Treatments were applied directly to mature seeds (M) prior to germination and early seedling growth. PC1 (56.96%) and PC2 (22.08%) capture strong transcriptomic shifts associated with seedling development from germination (G) through 7 and 10 days post-sowing (7D, 10D). Distinct clusters for each time point highlight dynamic transcriptomic reprogramming influenced by the seed-applied treatments. Each point represents a single biological replicate. All tissue types and treatment conditions are included. Developmental stages (S1, S2, M, G, 7D, 10D) are labelled for interpretability.*

*Transcriptomic analysis:* The transcriptomic analysis, based on RNA-Seq technology and performed on 144 wheat samples (Genotype 4), revealed a specific reprogramming of gene expression in seeds derived from treated mother plants, as well as in seeds subjected to coating treatments. A significant set of genes associated with defence responses was found to be differentially expressed. However, our results indicate that developmental stage and tissue type had a more pronounced impact on gene expression variation than the treatments themselves (Figure 1). This suggests that, while biocontrol treatments do elicit transcriptional responses (Figure 2), these effects may be modulated by the physiological context in which the treatment is applied.

Figure 2: Differentially Expressed Genes (DEGs) in Response to Treatments Across Wheat Grain Developmental Stages

Figure 2 : Gènes Différentiellement Exprimés (DEGs) en Réponse aux Traitements Selon les Stades de Développement du Grain de Blé

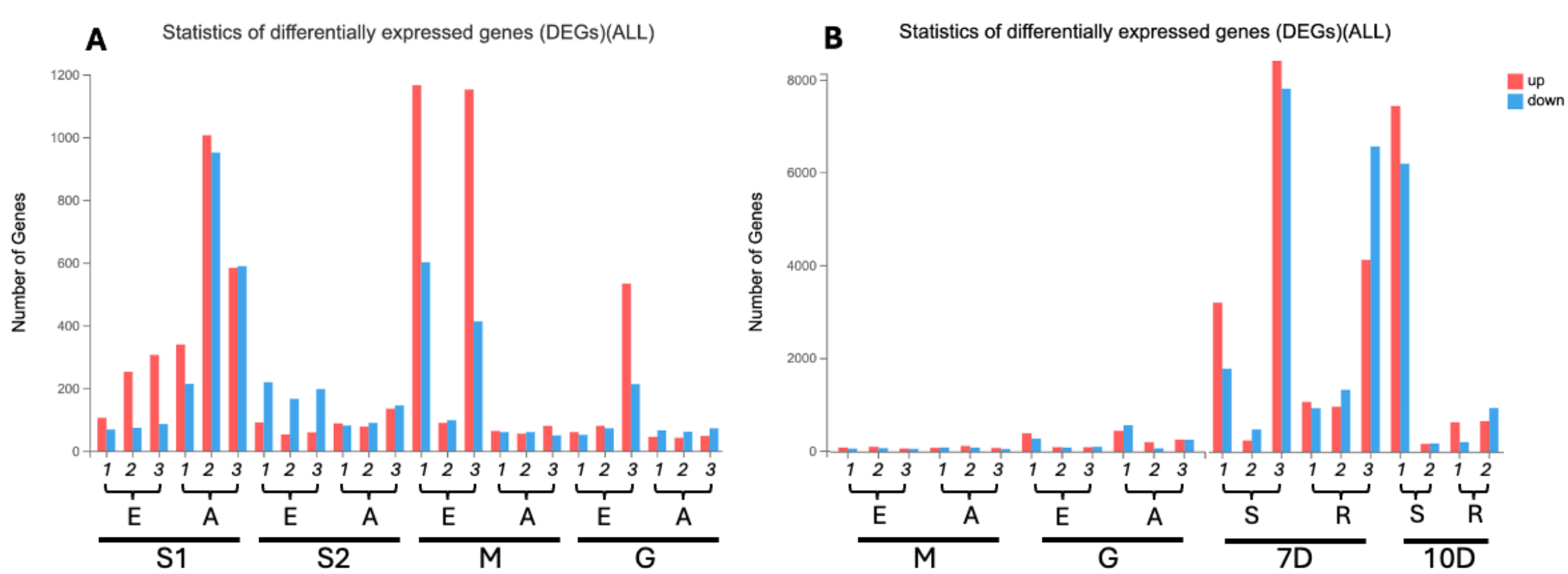

*Bar plots show the number of DEGs, either upregulated (red) or downregulated (blue), in response to treatments. In both panels, bars represent pairwise comparisons between the water-treated control and each treatment: [1] =* Pseudomonas chlororaphis *(BC, biocontrol agent), [2] = copper oxide (Co), and [3] = acibenzolar-S-methyl (ASM). Gene counts are shown on the y-axis, and conditions grouped by stage are displayed on the x-axis. Numbers below the x-axis indicate the treatments. For both panels, (A, left) and (B, right), significantly upregulated genes are shown in red and downregulated genes in blue.*

***(Panel A)*** *Pre-harvest treatment: Seeds were collected from mother plants treated during seed development. Tissues were analysed at three seed development stages: early maturation [S1], late* *maturation [S2], and mature dry seed [M], as well as at the germinated stage [G]. For S1, S2, M, and G, samples were separated into embryo [E] and endosperm plus bran [A].*
***(Panel B)*** *Post-harvest treatment: Seeds were treated by coating and analysed at germination [G] and during early seedling growth (7 days [7D] and 10 days [10D] post-sowing). For M and G, samples were separated into embryo [E] and endosperm plus bran [A]. For seedlings at 7D and 10D, root [R] and shoot [S] tissues were analysed separately to capture organ-specific transcriptome responses. Marked transcriptional changes were observed particularly at 7D and 10D, reflecting extensive transcriptome reprogramming during early seedling development.*

As shown in Figure 3, the expression of genes encoding pathogen recognition receptors, genes involved in plant–pathogen interactions, and components of the MAPK signalling pathway was significantly altered. These genes encode key proteins involved in the early detection of pathogens and the initiation of immune responses.

Figure 3: Enrichment Analysis of KEGG Pathways Among Differentially Expressed Genes in Embryos and Shoots Following Biocontrol Treatments
Figure 3 : Analyse d'Enrichissement des Voies KEGG Parmi les Gènes Différentiellement Exprimés dans les Embryons et les Parties Aériennes des Plantules en Réponse aux Traitements de Biocontrôle

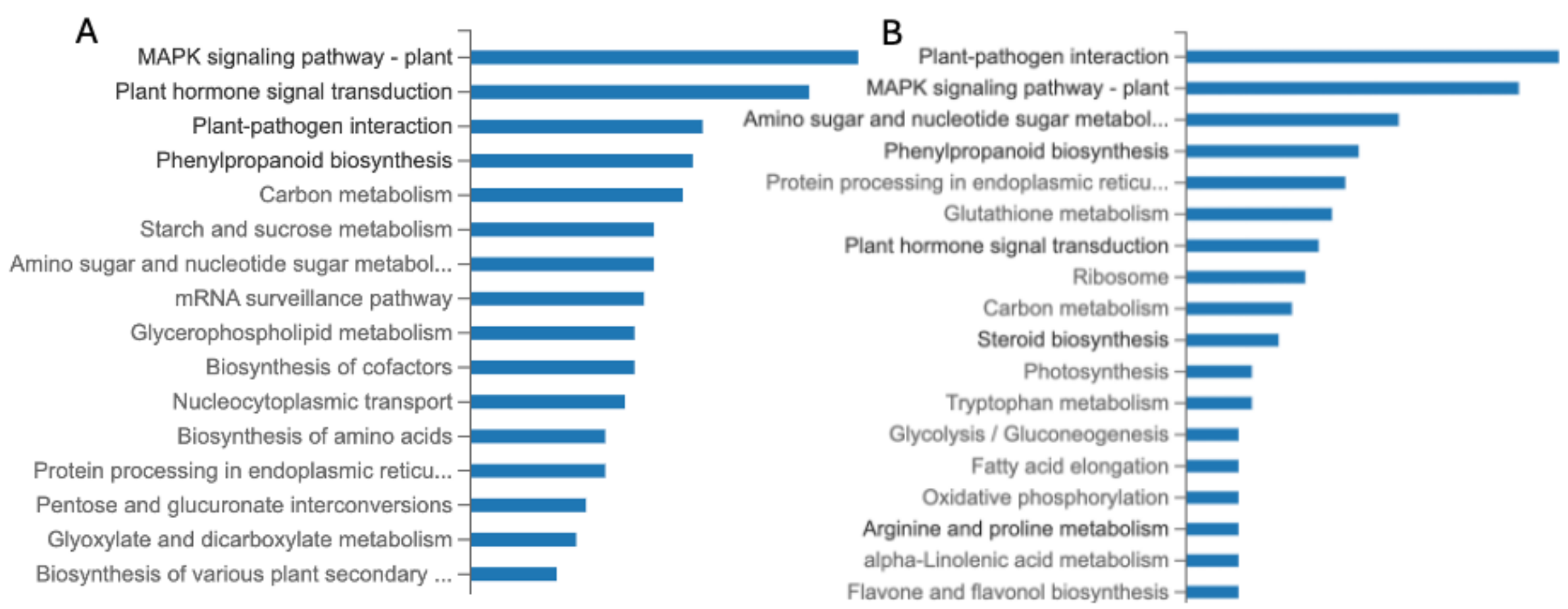

*Bar plots showing significantly enriched KEGG pathways (https://www.genome.jp/kegg/pathway.html) among differentially expressed genes (DEGs) identified in embryo (A) and shoot tissues (B) of wheat under biocontrol treatments with* Pseudomonas *chlororaphis strain (BC).* ***(Panel A)*** *Embryos were collected from seeds produced by mother plants sprayed with the biocontrol agent during seed development (pre-harvest treatment).* ***(Panel B)*** *Shoots were harvested from 7-day-old seedlings grown from seeds coated with the biocontrol agent (post-harvest seed treatment). Pathways related to plant defence and signalling, such as "Plant-pathogen interaction", "MAPK signalling pathway - plant", and "Plant hormone signal transduction" are among the most enriched, suggesting that both maternal and seed-applied treatments with* Pseudomonas chlororaphis *(BC, biocontrol agent) strongly reprogram gene expression toward enhanced immune and metabolic readiness. This analysis highlights distinct, tissue-specific transcriptomic reprogramming patterns associated with the mode and timing of biocontrol application.*

Furthermore, a notable activation of genes involved in plant hormone signal transduction was observed, indicating a coordinated mobilisation of systemic defence pathways. This regulation was accompanied by the upregulation of genes encoding enzymes involved in the biosynthesis of antimicrobial compounds, such as flavonoids and phenolic compounds, *via* the phenylpropanoid biosynthesis pathway. A clear distinction was observed between the effects of treatments applied to the mother plant and those applied directly to the seeds by coating. Treatment of the mother plant induced a broader and more intense transcriptional response, suggesting a form of immune priming, possibly even a memory, transmitted to the seeds, although this phenomenon remains to be confirmed. In contrast, seed coating triggered a more specific and moderate activation of defence-related genes, without signs of deleterious stress or disruption of basal metabolism (Figure 2). Tissue comparison also revealed transcriptomic specificity, with distinct expression profiles depending on the developmental stage. This highlights the pivotal role of phenology in determining sensitivity to biocontrol agents and the perception of environmental cues. In the absence of negative effects on germination, these findings confirm that biocontrol agents, when applied in these modalities, effectively activate plant defences while preserving seed quality and early seedling vigour. These results support the potential of such approaches in the development of integrated seed protection strategies, combining biological efficacy with respect for physiological processes.

*Metabolomics*: Metabolomic profiling, performed by liquid chromatography coupled with high-resolution tandem mass spectrometry (LC-MS/MS QTOF) on 144 samples of the Genotype 4, complemented transcriptomic observations by providing a functional view of the biochemical changes induced by biocontrol treatments. The resulting metabolic profiles revealed significant variations in the repertoire of specialised metabolites, some of which were clearly associated with plant defence mechanisms. In particular, a marked accumulation of compounds with antifungal and antibacterial potential was observed in seeds from treated mother plants, notably flavonoids and other phenolic derivatives. These molecules play a central role in protecting against a broad range of pathogens, and their biosynthesis is commonly triggered during the activation of plant immune responses, notably through the phenylpropanoid pathway or the production of specific dipeptides. Moreover, metabolites involved in hormone-mediated defence signalling, such as intermediates of the shikimate pathway, were also found at elevated levels in treated seed samples. These results corroborate transcriptomic data and suggest a coordinated functional activation of defence pathways at both transcriptional and metabolic levels. Comparison between the two treatment modalities, application to the mother plant or directly to the seed by coating, indicated that the impact on the seed metabolome was considerably more pronounced when the treatment was applied to the mother plant. This approach resulted in a greater diversification of detected defence-related compounds, which may reflect a more comprehensive metabolic priming of the seeds during their development. This phenomenon suggests that such seeds may possess enhanced capacities to withstand biotic stress during germination and early growth stages. Overall, these findings demonstrate that the treatments used durably modify the metabolic profile of wheat seeds by promoting the accumulation of compounds associated with environmental and biotic responses, without impairing germination capacity. This reinforced metabolic response likely represents one of the key mechanisms supporting enhanced protection against biotic stress in both seeds and seedlings.

**Discussion**

Our results demonstrate that the application of a biocontrol agent such as *Pseudomonas chlororaphis* to wheat, either *via* foliar treatment of the mother plant or seed coating, triggers a substantial defence response without impairing seed germination or early seedling vigour. Transcriptomic analyses revealed the coordinated activation of genes involved in plant-pathogen interactions, MAPK signalling cascades, and hormone signal transduction pathways. These findings are consistent with previous studies which demonstrated defence activation by bacterial cyclodipeptides (Song et al., 2017), and they further confirm the central role of hormonal pathways in establishing robust immune responses in seeds and seedlings. In parallel, the metabolomic changes observed in seeds treated with a biocontrol agent align with previous studies reporting the accumulation of specific metabolites, namely flavonoids and phenolics, well known for their antimicrobial activities (Santoyo et al., 2024; Hu et al., 2025). The convergence of transcriptomic and metabolomic data supports the view that responses induced are not merely indicative of activation but constitute a coordinated and functional defence programme. Furthermore, these treatments may influence the assembly and function of the seed- and seedling-associated microbiota during early development, contributing to a broader and more durable protective effect. This interpretation is consistent with recent meta-analyses identifying microbiota modulation as a promising agroecological practice (Lamichhane et al., 2022). Critically, these findings must be framed within the context of the seed as a biologically and ecologically unique organ. As recently emphasised (Hubert et al., 2024), seeds are not passive structures but deploy a suite of constitutive defence mechanisms, including physical barriers, biochemical deterrents, and the controlled release of antimicrobial exudates into the spermosphere upon imbibition. These defences are tightly regulated throughout developmental stages and enable the seed to withstand prolonged exposure to pathogens during dormancy. This perspective is reinforced by the work of Dalling et al. (2020), who extended plant defence theory to include seeds, demonstrating that seeds possess distinctive defence strategies shaped by the dual pressures of survival and effective germination. These strategies often involve trade-offs with dispersal and seedling performance and highlight seeds as autonomous actors in biotic interactions. Moreover, the interaction between seeds and microbial communities (both mutualistic and antagonistic) emerges as a key determinant of seed viability and seedling success. Recent advances also underscore the singularity of the seed microbiome itself. Abdelfattah et al. (2023) propose that seeds act as a microbial inheritance bottleneck, playing a pivotal role in shaping the plant microbiome across generations. Their opinion piece highlights three critical phases in microbial inheritance, (i) plant-to-seed transfer, (ii) the dormancy period, and (iii) seed-to-seedling transition, each governed by distinct ecological and physiological processes. These findings reinforce the idea that seed microbiota are not merely subsets of the plant microbiome, but structured communities with specific transmission and functional roles. In this light, biocontrol strategies targeting seeds must consider not only plant–pathogen interactions, but also the seed's native microbiome as both a mediator and a partner in defence and development. In our study, maternal treatment induced stronger and more diverse transcriptional and metabolic responses compared to seed coating. This may reflect a more anticipatory defence programming during seed development, potentially resulting in enhanced preparedness of seeds to respond to early biotic challenges. Such maternal imprinting could resemble a form of immune memory, though this remains to be validated mechanistically. To ensure the durable integration of such seed-targeted approaches in agriculture, they must be embedded within an integrated pest management framework. Combining microbial biocontrol with agroecological practices, such as crop rotation, varietal mixtures, and reduced chemical inputs, can lower pathogen pressure while delaying the emergence of resistance. However, the observed genotypic and species-level variation in responses also underscores the need for tailored strategies adapted to the specific physiological and microbial features of the seed. Finally, the development of efficient microbial seed treatments will require a deeper understanding of seed-specific physiology, defence programmes, and microbiome dynamics. Future research should explore the stability, heritability, and ecological consequences of biocontrol-induced responses across generations and systematically assess the impact of these treatments on the seed-associated microbiota. Only through such integrated research efforts will it be possible to design robust, effective, and ecologically sound seed enhancement strategies for sustainable agriculture.

**Conclusion/perspectives**

Our findings demonstrate that biocontrol agents, when applied either to the mother plant or directly to seeds, can effectively activate plant defence responses without compromising seed quality. This approach presents a promising avenue for enhancing the sustainability of seed production systems. Beyond these results, several key research challenges remain to be taken on. These include investigations into the stability and potential transgenerational transmission of biocontrol-induced responses, as well as assessing the impact of treatments on the seed-specific microbiome. Addressing these questions will be critical for the development of more targeted and durable bio-inspired protection strategies, aligned with the goals of agroecological transition.

**Acknowledgements**
The authors gratefully acknowledge the Biocontrol Consortium and PlantAlliance for their financial support of the SeedBioProtect project. The authors also wish to thank the Saclay Plant Sciences (SPS) LabEx for supporting IJPB (ANR-10-LABX-0040-SPS). In addition, the authors thank RAGT Seeds for providing the wheat genotypes and Koppert for supplying the biocontrol solution.